\documentclass[conference]{IEEEtran}
\usepackage[noadjust]{cite}

\ifCLASSINFOpdf
\else
\fi
\usepackage{amsmath}
\usepackage{amssymb}

\renewcommand{\IEEEQED}{\IEEEQEDopen}

\newcommand{\docVersion}{0.5.1}
\newcommand{\docBuildNumber}{?}
\InputIfFileExists{bn.aux}{
 \typeout{Build number set!}
}{}

\usepackage{color}
\newcommand{\omitstyle}{\color[rgb]{0.25,0,0.75}}
\newcommand{\omitted}[1]{{\omitstyle#1}}
\InputIfFileExists{option.aux}{%
 \typeout{Compiling options activated!}%
}{}

\newtheorem{theorem}{Theorem}[section]
\newtheorem{corollary}[theorem]{Corollary}

\newtheorem{proposition}[theorem]{Proposition}
\newtheorem{definition}[theorem]{Definition}

\newtheorem{remark}[theorem]{Remark}
\newtheorem{trick}[theorem]{Trick}

\newcommand{\eqdef}{:=}

\newcommand{\tovar}[1]{\buildrel#1\over\longrightarrow}

\newcommand{\fromOneTo}[1]{[#1]}
\newcommand{\real}{\mathbf{R}}
\newcommand{\realOf}[1]{\real_{#1}}
\newcommand{\borel}{\mathfrak{B}}

\newcommand{\emptySet}{\varnothing}
\newcommand{\setComplement}{\mathsf{c}}
\newcommand{\powerSet}{\mathcal{P}}
\newcommand{\closure}[1]{\overline{#1}}
\newcommand{\diff}{\mathrm{d}}
\newcommand{\kp}{\times}
\DeclareMathOperator{\expect}{E}
\DeclareMathOperator{\statDistance}{d}

\newcommand{\halfProjLine}{\mathbb{P}}
\newcommand{\distHPL}{d_{\mathbb{P}}}

\newcommand{\uniDistribution}[1]{\mathrm{U}_{#1}}

\DeclareMathOperator{\littleO}{o}

\newcommand{\term}[1]{\emph{#1}}

\DeclareMathOperator{\entropyRate}{\mathcal{H}}
\DeclareMathOperator{\infEntropyRate}{\underline{\entropyRate}}
\DeclareMathOperator*{\pliminf}{p-lim\,inf}

\begin{document}
%
\title{Beyond Countable Alphabets: An Extension of the Information-Spectrum Approach}



%
\author{
 \IEEEauthorblockN{%
  Shengtian Yang\IEEEauthorrefmark{1}\IEEEauthorrefmark{2},
  Thomas Honold\IEEEauthorrefmark{3}\IEEEauthorrefmark{2},
  and Zhaoyang Zhang\IEEEauthorrefmark{3}\IEEEauthorrefmark{2}%
 }
 \IEEEauthorblockA{%
  \IEEEauthorrefmark{1}%
  School of Information and Electronic Engineering,
  Zhejiang Gongshang University, Hangzhou 310018, China%
 }
 \IEEEauthorblockA{%
  \IEEEauthorrefmark{2}%
  Zhejiang Provincial Key Laboratory of Information Processing,
  Communication and Networking, Hangzhou 310027, China%
 }
 \IEEEauthorblockA{%
  \IEEEauthorrefmark{3}%
  College of Information Science and Electronic Engineering,
  Zhejiang University, Hangzhou 310027, China%
 }
 \IEEEauthorblockA{%
  Email: \texttt{yangst@codlab.net, honold@zju.edu.cn, ning\_ming@zju.edu.cn}%
 }
}


\maketitle

\begin{abstract}
A general approach is established for deriving one-shot performance
bounds for information-theoretic problems on general alphabets beyond
countable alphabets.
It is mainly based on the quantization idea and a novel form of
``likelihood ratio''.
As an example, one-shot lower and upper bounds for random number
generation from correlated sources on general alphabets are derived.
\end{abstract}


%
\IEEEpeerreviewmaketitle

\section{Introduction}
The information-spectrum approach, since its introduction by Han and
Verd\'{u} \cite{han_approximation_1993}, has become one of the most
important tools of information theory.
Similar to Shannon's classical approach (see e.g.,
\cite{cover_elements_2006}), the information-spectrum approach is
mainly confined to countable alphabets, though in some cases, its
extension to continuous alphabets is direct.
The difficulty comes from two aspects:

1) It is difficult to formulate information measures and performance
bounds for arbitrary alphabets in a unified way.
The Radon-Nikodym derivative seems a good candidate, or a good basis
for defining information measures, but it exists only if the
absolute-continuity condition is satisfied.

2) It is difficult to extend certain useful proof techniques from
finite alphabets to general alphabets, and even to countably infinite
alphabets.
For example, it is not easy to construct an analog of the random-bin
map when the domain is uncountable, and it is also difficult to
generalize structured random coding techniques (such as random
matrices) to the case of infinite alphabets.

This paper will partly solve this problem by providing a general
approach for one-shot performance bounds, an important part of the
information-spectrum approach.
Our main approach is based on quantization, which effectively
overcomes the second difficulty.
Given any problem on infinite alphabets, we take the following steps:
1) modifying the problem by quantization so that the modified version
can be described on finite alphabets and hence has already a solution;
2) converting the solution of the modified problem into a solution of
the original problem;
and 3) repeating the first and second steps with a sequence of
quantizations with increasing resolution to get the asymptotically
optimal solution.
This idea may look simple, but is difficult to be developed into a
general approach.
In the rest of this paper, we will give the basic results of this
approach and then illustrate this approach by an example.
We will also introduce a novel form of ``likelihood ratio'', a
generalization of the Radon-Nikodym derivative that perfectly solves
the first difficulty.

We close this section with some notations used throughout this paper.
The field of real numbers is denoted by $\real$, and the set of
integers from $1$ to $n$ is denoted by $\fromOneTo{n}$.
When performing probabilistic analysis, all objects of study are
related to a basic probability space $(\Omega,\mathfrak{A},P)$ with
$\mathfrak{A}$ a $\sigma$-algebra in $\Omega$ and $P$ a probability
measure on $(\Omega,\mathfrak{A})$.
A random element in a measurable space $(\mathcal{X},\mathfrak{X})$ is
a measurable mapping from $\Omega$ into $\mathcal{X}$.
For probability measures $\mu$ and $\nu$ on
$(\mathcal{X},\mathfrak{X})$, the \term{statistical distance} between
$\mu$ and $\nu$ is
\[
\statDistance(\mu,\nu)
\eqdef \sup_{C\in\mathfrak{X}} |\mu(C)-\nu(C)|.
\]

Measure-theoretic methods will be used in this paper.
Readers not familiar with measure theory are referred to
\cite{cohn_measure_2013,durrett_probability:_2010,kallenberg_foundations_1997}.
All measures considered in this paper are finite.
For a topological space $S$, its Borel $\sigma$-algebra generated by
the topology in $S$ is denoted by $\borel(S)$.
The integral of a real-valued, measurable function $f$ on some measure
space $(\mathcal{X},\mathfrak{X},\lambda)$ is denoted by
$\lambda f=\lambda(f)$.
When $\lambda$ is a probability measure, we write
$\expect_\lambda f=\expect_\lambda(f)$ in place of $\lambda f$ and
write $\expect f$ if $\lambda=P$.
If $\lambda|f|<+\infty$, then $\lambda f$ induces a finite signed
measure $(f\cdot\lambda)(A)\eqdef\lambda(f1_A)$ on
$(\mathcal{X},\mathfrak{X})$.
Given a sub $\sigma$-algebra $\mathfrak{F}$ of $\mathfrak{X}$, we
denote by $\lambda^{\mathfrak{F}} f$ the density function
$\diff(f\cdot\lambda)|_{\mathfrak{F}}/\diff\lambda|_{\mathfrak{F}}$,
which is called the conditional expectation of $f$ with respect to
$\mathfrak{F}$ if $\lambda$ is a probability measure.
In this case, we wirte $\expect_\lambda^\mathfrak{F} f$ in place of
$\lambda^{\mathfrak{F}} f$.
For $A\in\mathfrak{X}$,
$\lambda^{\mathfrak{F}}(A)\eqdef\lambda^{\mathfrak{F}}1_A$ defines a
kernel from $(\mathcal{X},\mathfrak{F})$ to
$(\mathcal{X},\mathfrak{X})$.
For a kernel $\mu$ from $(\mathcal{X},\mathfrak{X})$ to
$(\mathcal{Y},\mathfrak{Y})$, we denote by $\mu^B$ the real-valued map
$x\mapsto\mu(x,B)$ for some fixed $B\in\mathfrak{Y}$, and by $\mu_x$
the measure on $(\mathcal{Y},\mathfrak{Y})$ for some fixed
$x\in\mathcal{X}$.
Thus $\lambda(\mu^B)$, as a function of $B$, becomes a measure on
$(\mathcal{Y},\mathfrak{Y})$, and is usually written as
$\lambda(\mu)$.
The extended kernel $\overline{\mu}$ of $\mu$ is a kernel from
$(\mathcal{X},\mathfrak{X})$ to
$(\mathcal{X}\times\mathcal{Y},\mathfrak{X}\times\mathfrak{Y})$ given
by $(x,C)\mapsto\mu(x,C_x)$ with $C_x=\{y:(x,y)\in C\}$.
The product $\lambda\kp\mu\eqdef\lambda(\overline{\mu})$ is a measure
on $(\mathcal{X}\times\mathcal{Y},\mathfrak{X}\times\mathfrak{Y})$,
and we simply write $\lambda\mu$ when there is no possible ambiguity.
Note that $\lambda\mu$ coincides with the product-measure notion when
$\mu$ reduces to a measure on $(\mathcal{Y},\mathfrak{Y})$.

\section{The Quantization Approach}\label{sec:Quantization}

In this section, we will establish the basic results of the
quantization approach.
Because of the space limitation, most simple proofs are omitted.
The readers are referred to \cite{yang_beyond_2016} for omitted
proofs.

We first give an overview of the main tricks of our approach.

\begin{trick}[Quantization]
Given a measurable space $(\mathcal{X},\mathfrak{X}$), a finite
quantization of $\mathcal{X}$ can be characterized by a finite sub
$\sigma$-algebra $\mathfrak{F}$ of $\mathfrak{X}$, which induces a
natural projection $\pi_{\mathfrak{F}}$ from $\mathcal{X}$ to
$\mathrm{atoms}(\mathfrak{F})$ given by $x\mapsto z$ such that
$x\in z$, where $\mathrm{atoms}(\mathfrak{F})$ is the set of all
elements in $\mathfrak{F}$ that cannot be decomposed into smaller
pieces that are also in $\mathfrak{F}$.
In fact, $\mathrm{atoms}(\mathfrak{F})$ forms a finite partition of
$\mathcal{X}$.
\end{trick}

\begin{trick}[Approximation by Theorems~\ref{th:FiniteApproximation},
\ref{th:ConditionalProbability}, \ref{th:FiniteApproximation.X}, and
Corollary~\ref{co:FiniteApproximation}]
\label{tr:Approximation}
Let $f=(f_i)_{i=1}^\ell$ be a family of real-valued integrable
functions on the measure space
$(\mathcal{X}_1\times\mathcal{X}_2,
 \mathfrak{X}_1\times\mathfrak{X}_2,\mu)$.
Let $\epsilon>0$.
By Theorem~\ref{th:FiniteApproximation}, for each
$i\in\fromOneTo{\ell}$, there is a finite sub $\sigma$-algebra
$\mathfrak{C}_{i,j}$ of $\mathfrak{X}_j$ for each
$j\in\fromOneTo{2}$ such that
\[
\mu|\mu^{\mathfrak{F}}f_i-f_i|<\epsilon
\]
for every $\sigma$-algebra $\mathfrak{F}$ satisfying
$\mathfrak{C}_{i,1}\times\mathfrak{C}_{i,2}\subseteq\mathfrak{F}
 \subseteq\mathfrak{X}_1\times\mathfrak{X}_2$.
Taking
\[
\mathfrak{D}_j
= \sigma\left(\bigcup_{i\in\fromOneTo{\ell}}\mathfrak{C}_{i,j}\right)
\]
and $\mathfrak{F} = \mathfrak{D}_1\times \mathfrak{D}_2$, we thus have
\[
\mu|\mu^{\mathfrak{F}}f_i-f_i|<\epsilon
\]
for all $i\in\fromOneTo{\ell}$.
In the same vein and by Corollary~\ref{co:FiniteApproximation}, we can
show that, for any $\nu\ll\mu$, there is a finite sub $\sigma$-algebra
$\mathfrak{F}$ of $\mathfrak{X}_1\times\mathfrak{X}_2$ such that
\[
\nu\{|\mu^{\mathfrak{F}}f_i-f_i|\ge\epsilon\}<\epsilon
\]
for all $i\in\fromOneTo{\ell}$.
Then we can find a sequence $(\mathfrak{F}_n)_{n=1}^\infty$ of finite
sub $\sigma$-algebras so that each sequence
$g_i^{(n)}=\mu^{\mathfrak{F}_n}f_i$ converges in $\nu$-measure to
$f_i$, or further,
\[
\|(g_i^{(n)})_{i=1}^\ell-(f_i)_{i=1}^\ell\|_p
\tovar{\nu} 0
\]
for any $p$-norm of $\real^\ell$ with $p\ge 1$, where
$\buildrel\nu\over\to$ is the shorthand of convergence in
$\nu$-measure.
Similar tricks with Theorems~\ref{th:ConditionalProbability}
and \ref{th:FiniteApproximation.X} also work for statistical distances
and likelihood ratios (Trick~\ref{tr:LikelihoodRatio}).
\end{trick}

\begin{trick}[Handling likelihood ratios by
Theorems~\ref{th:FiniteApproximation.X}]
\label{tr:LikelihoodRatio}
Likelihood ratios may be the objects most often occurring in a one-shot
performance bound.
Let $\mu$ and $\nu$ be two measures on
$(\mathcal{X}\times\mathcal{Y},\mathfrak{X}\times\mathfrak{Y})$.
The likelihood ratio of $\mu$ to $\nu$ is
\[
[\diff\mu:\diff\nu]
= [\diff\mu/\diff(\mu+\nu):\diff\nu/\diff(\mu+\nu)]
\quad \text{(Definition~\ref{de:LikelihoodRatio})},
\]
a $(\halfProjLine,\borel(\halfProjLine))$-valued measurable function
on $(\mathcal{X}\times\mathcal{Y},\mathfrak{X}\times\mathfrak{Y})$,
where $\halfProjLine$ is the half projective line defined by
Definitions~\ref{de:HPL} and \ref{de:HPLMetricAndOrder}.
For any $\xi\ll\mu+\nu$, we can find a sequence
$(\mathfrak{F}_n)_{n=1}^\infty$ of finite sub $\sigma$-algebras of
$\mathfrak{X}\times\mathfrak{Y}$ such that
\begin{IEEEeqnarray*}{r}
\distHPL([\diff\mu|_{\mathfrak{F}_n}:\diff\nu|_{\mathfrak{F}_n}],
 [\diff\mu:\diff\nu])
\tovar{\xi} 0\qquad\qquad\qquad\\
\text{(Theorem~\ref{th:FiniteApproximation.X}
and Trick~\ref{tr:Approximation})}.
\end{IEEEeqnarray*}
\end{trick}

Having introduced the main tricks, we proceed to introduce the details
of the quantization approach.

\begin{theorem}\label{th:FiniteApproximation}
Let $f$ be a real-valued integrable function on the measure space
$(\mathcal{X}\times\mathcal{Y},
 \mathfrak{X}\times\mathfrak{Y},\mu)$.
Then for any $\epsilon>0$, there is a finite sub $\sigma$-algebra
$\mathfrak{C}$ of $\mathfrak{X}$ and a finite sub $\sigma$-algebra
$\mathfrak{D}$ of $\mathfrak{Y}$ such that
\begin{equation}
\mu|\mu^{\mathfrak{F}}f-f|<\epsilon \label{eq:FiniteApproximation.1}
\end{equation}
for every $\sigma$-algebra $\mathfrak{F}$ satisfying
$\mathfrak{C}\times\mathfrak{D}\subseteq\mathfrak{F}
 \subseteq\mathfrak{X}\times\mathfrak{Y}$.
\end{theorem}

\begin{IEEEproof}
We say that a real-valued integrable function $f$ on
$(\mathcal{X}\times\mathcal{Y},\mathfrak{X}\times\mathfrak{Y},\mu)$
can be finitely approximated if for any $\epsilon>0$, there are a
finite sub $\sigma$-algebra $\mathfrak{C}$ of $\mathfrak{X}$ and a
finite sub $\sigma$-algebra $\mathfrak{D}$ of $\mathfrak{Y}$ such that
$\mu|\mu^{\mathfrak{F}}f-f|<\epsilon$ for every $\sigma$-algebra
$\mathfrak{F}$ satisfying
$\mathfrak{C}\times\mathfrak{D}\subseteq\mathfrak{F}
 \subseteq\mathfrak{X}\times\mathfrak{Y}$.
We define
\[
\mathcal{H}
= \{f: \text{$f$ is integrable and can be finitely
 approximated}\}
\]
and $\mathcal{A}=\{C\times D: C\in\mathfrak{X},D\in\mathfrak{Y}\}$.
It is clear that $\mathcal{A}$ is a $\pi$-system containing
$\mathcal{X}\times\mathcal{Y}$, and we have:

(a) If $A=C\times D\in\mathcal{A}$, then $1_A\in\mathcal{H}$ with
$\mathfrak{C}=\{\emptySet,C,C^\setComplement,\mathcal{X}\}$ and
$\mathfrak{D}=\{\emptySet,D,D^\setComplement,\mathcal{Y}\}$.

(b) If $f,g\in\mathcal{H}$, then there are finite sub
$\sigma$-algebras $\mathfrak{C}'$, $\mathfrak{C}''$ of $\mathfrak{X}$
and $\mathfrak{D}'$, $\mathfrak{D}''$ of $\mathfrak{Y}$ such that
$\mu|\mu^{\mathfrak{F}'}f-f|<\epsilon/2$ and
$\mu|\mu^{\mathfrak{F}''}g-g|<\epsilon/2$ for every $\sigma$-algebra
$\mathfrak{C}'\times\mathfrak{D}'\subseteq\mathfrak{F}'
 \subseteq\mathfrak{X}\times\mathfrak{Y}$
and every $\sigma$-algebra
$\mathfrak{C}''\times\mathfrak{D}''\subseteq\mathfrak{F}''
 \subseteq\mathfrak{X}\times\mathfrak{Y}$,
respectively.
Then
\[
\mu|\mu^{\mathfrak{F}}(f+g)-(f+g)|
\le \mu|\mu^{\mathfrak{F}}f-f| + \mu|\mu^{\mathfrak{F}}g-g|
< \epsilon
\]
for every $\sigma$-algebra
$\mathfrak{C}\times\mathfrak{D}\subseteq\mathfrak{F}
 \subseteq\mathfrak{X}\times\mathfrak{Y}$
with $\mathfrak{C}=\sigma(\mathfrak{C}'\cup\mathfrak{C}'')$ and
$\mathfrak{D}=\sigma(\mathfrak{D}'\cup\mathfrak{D}'')$.
In other words, $f+g\in\mathcal{H}$.
In a similar way, we can show that $cf\in\mathcal{H}$ for $c\in\real$.

(c) If $f_n\in\mathcal{H}$ converges everywhere to an integrable
function $g$ (including all bounded functions for finite $\mu$) with
$|f_n|\le|g|$, then by the dominated convergence theorem,
$\mu|f_N-g|<\epsilon/4$ for some large integer $N$.
Furthermore, since $f_N\in\mathcal{H}$, there are finite
$\sigma$-algebras $\mathfrak{C}$ of $\mathfrak{X}$ and $\mathfrak{D}$
of $\mathfrak{Y}$ such that
$\mu|\mu^{\mathfrak{F}}f_N-f_N|<\epsilon/2$ for every $\sigma$-algebra
$\mathfrak{C}\times\mathfrak{D}\subseteq\mathfrak{F}
 \subseteq\mathfrak{X}\times\mathfrak{Y}$,
so that
\begin{IEEEeqnarray*}{rCl}
\mu|\mu^{\mathfrak{F}}g-g|
&\le &\mu|\mu^{\mathfrak{F}}g-\mu^{\mathfrak{F}}f_N|
 + \mu|\mu^{\mathfrak{F}}f_N-f_N| + \mu|f_N-g|\\
&\le &\mu|g-f_N| + \mu|\mu^{\mathfrak{F}}f_N-f_N| + \mu|f_N-g|\\
&= &2\mu|f_N-g| + \mu|\mu^{\mathfrak{F}}f_N-f_N|
< \epsilon
\end{IEEEeqnarray*}
for every $\sigma$-algebra
$\mathfrak{C}\times\mathfrak{D}\subseteq\mathfrak{F}
 \subseteq\mathfrak{X}\times\mathfrak{Y}$.
Therefore, $g\in\mathcal{H}$.

By the monotone class theorem for functions
(\cite[Theorem~6.1.3]{durrett_probability:_2010}) with properties
(a)--(c), we conclude that $\mathcal{H}$ contains all bounded
functions measurable with respect to
$\sigma(\mathcal{A})=\mathfrak{X}\times\mathfrak{Y}$.
Again by (c) with $f_n(p)=g(p)1\{g(p)\le n\}$ for arbitrary integrable
$g$, it is easy to see that $\mathcal{H}$ contains all integrable
functions.
\end{IEEEproof}


\begin{corollary}\label{co:FiniteApproximation}
Let $f$ be a real-valued integrable function on the measure space
$(\mathcal{X}\times\mathcal{Y},
 \mathfrak{X}\times\mathfrak{Y},\mu)$.
Let $\nu$ be a measure such that $\nu\ll\mu$.
Then for any $\epsilon>0$, there is a finite sub $\sigma$-algebra
$\mathfrak{C}$ of $\mathfrak{X}$ and a finite sub $\sigma$-algebra
$\mathfrak{D}$ of $\mathfrak{Y}$ such that
\[
\nu\{|\mu^{\mathfrak{F}}f-f|\ge\epsilon\}<\epsilon
\]
for every $\sigma$-algebra $\mathfrak{F}$ satisfying
$\mathfrak{C}\times\mathfrak{D}\subseteq\mathfrak{F}
 \subseteq\mathfrak{X}\times\mathfrak{Y}$.
\end{corollary}

\begin{theorem}\label{th:ConditionalProbability}
Let $\mu$ be a probability kernel from
$(\mathcal{X},\mathfrak{X},\lambda)$ to
$(\mathcal{Y},\powerSet(\mathcal{Y}))$ with $\mathcal{Y}$ at most
countable.
Then for any $\epsilon>0$, there is a finite sub $\sigma$-algebra
$\mathfrak{C}$ of $\mathfrak{X}$ such that
\[
\expect_\lambda\statDistance(\expect_\lambda^\mathfrak{F}\mu,\mu)
< \epsilon
\]
for every $\sigma$-algebra $\mathfrak{F}$ satisfying
$\mathfrak{C}\subseteq\mathfrak{F}\subseteq\mathfrak{X}$.
\end{theorem}

A performance bound obtained by the information-spectrum approach is
often expressed in terms of some kind of random likelihood ratios, it
is thus necessary to understand the notions of likelihood ratios well.
Usually, a likelihood ratio is expressed as a Radon-Nikodym derivative
of two probability measures.
Sometimes, however, we will encounter a more complicated form of
likelihood ratios, a Radon-Nikodym derivative of two probability
kernels.
In the discrete case, it can be written as
$P_{Y\mid X}(y\mid x)/P_{\hat{Y}\mid X}(y\mid x)$, and we have
\[
\frac{P_{Y\mid X}(y\mid x)}{P_{\hat{Y}\mid X}(y\mid x)}
= \frac{P_{XY}(x,y)}{P_{X\hat{Y}}(x,y)}.
\]
A generalization of this identity is given as follows.

\begin{theorem}\label{th:KernelRatioToMeasureRatio}
Let $\mu$ and $\nu$ be two kernels from
$(\mathcal{X},\mathfrak{X},\lambda)$ to $(\mathcal{Y},\mathfrak{Y})$
such that $\mu_x\ll\nu_x$ for $\lambda$-almost every $x$ in
$\mathcal{X}$.
Then $\lambda\mu\ll\lambda\nu$ and for $\lambda$-almost every $x$,
\[
\frac{\diff(\lambda\mu)}{\diff(\lambda\nu)}(x,y)
=\frac{\diff\mu_x}{\diff\nu_x}(y)
\]
$\nu_x$-almost everywhere.
If $\diff\mu_x/\diff\nu_x$ has a version, say $f(x,y)$, that is
measurable with respect to $\mathfrak{X}\times\mathfrak{Y}$, then
$f=\diff(\lambda\mu)/\diff(\lambda\nu)$ $\lambda\nu$-almost
everywhere.
\end{theorem}

This theorem tells us that in general cases we need to use the form
$\diff(\lambda\mu)/\diff(\lambda\nu)$ in place of
$\diff\mu_x/\diff\nu_x$ because the former is always measurable with
respect to $\mathfrak{X}\times\mathfrak{Y}$.
A useful consequence of Theorem~\ref{th:KernelRatioToMeasureRatio} is:

\begin{corollary}\label{co:StatisticalDistance}
Let $\mu$ and $\nu$ be two probability kernels from the probability
space $(\mathcal{X},\mathfrak{X},\lambda)$ to
$(\mathcal{Y},\mathfrak{Y})$.
Then
\(
\statDistance(\lambda\mu,\lambda\nu)
= \expect_\lambda\statDistance(\mu_x,\nu_x).
\)
\end{corollary}

The Radon-Nikodym derivative $\diff\mu/\diff\nu$ cannot handle all
cases of likelihood ratios, because it does not exist if $\mu$ is not
absolutely continuous with respect to $\nu$.
In this case, we need a more general form of likelihood ratios based
on the approach of \cite{halmos_application_1949}.

Recall that the (real) projective line is defined as the set of
lines through the origin in the affine plane $\real^2$, and points
$\real(x,y)$ of the projective line are written as $(x:y)$
(homogeneous coordinates), reflecting the fact that
$\real(x,y)=\real(z,w)$ iff $x/y = z/w$ (for $y,w\ne 0$).
In analogy, we define $\halfProjLine$, the \emph{nonnegative part} of
the projective line, as follows:

\begin{definition}\label{de:HPL}
A pair $(r,s)\in\real^2$ is said to be admissible if
$(r,s)\in\realOf{\ge 0}^2\setminus\{(0,0)\}$.
The set $\realOf{>0}(r,s)\eqdef\{t(r,s):t>0\}$ forms a ray through the
origin in $\realOf{\ge 0}^2$ iff $(r,s)$ is admissible.
The \term{half projective line} $\halfProjLine$ is defined as the set
of rays through the origin in $\realOf{\ge 0}^2$, and points
$\realOf{>0}(r,s)$ of $\halfProjLine$ are written as $[r:s]$, or
simply as $r/s$ when $s\ne 0$ and there is no possible ambiguity.
The natural projection
$\pi_\halfProjLine:\realOf{\ge 0}^2\setminus\{(0,0)\}\to\halfProjLine$
given by $(x,y)\mapsto[x:y]$ thus induces a quotient topology in
$\halfProjLine$, so that $\pi_\halfProjLine$ becomes a quotient map
and is measurable with respect to the corresponding Borel
$\sigma$-algebras.
\end{definition}

Since the map
$\rho:\realOf{\ge 0}^2\setminus\{(0,0)\}\to[0,1]$ given by
\(
(x,y) \mapsto x/(x+y)
\)
is a quotient map, it follows from
\cite[Corollary~22.3]{munkres_topology_2000} that $\rho$ induces a
homeomorphism $\kappa:\halfProjLine\to[0,1]$ given by
\[
[x:y] \mapsto \frac{x}{x+y},
\]
which further induces a metric and an order on $\halfProjLine$.

\begin{definition}\label{de:HPLMetricAndOrder}
The metric $\distHPL$ on $\halfProjLine$ is defined by
\begin{IEEEeqnarray*}{rCl}
\distHPL([r_1:s_1],[r_2:s_2])
&\eqdef &|\kappa([r_1:s_1])-\kappa([r_2:s_2])|\\
&= &\left|\frac{r_1}{r_1+s_1}-\frac{r_2}{r_2+s_2}\right|.
\end{IEEEeqnarray*}
The order $\le$ of $\halfProjLine$ is defined by
\begin{IEEEeqnarray*}{rCl}
[r_1:s_1] \le [r_2:s_2]
&\Leftrightarrow &\kappa([r_1:s_1]) \le \kappa([r_2:s_2])\\
&\Leftrightarrow &r_1s_2-r_2s_1 \le 0.
\label{eq:HPLOrder}
\end{IEEEeqnarray*}
\end{definition}

It is clear that $\halfProjLine$ with metric $\distHPL$ is a complete
separable metric space.

For any real-valued functions $f$ and $g$ on $\mathcal{X}$, if
$(f(x),g(x))$ is admissible for all $x\in\mathcal{X}$, then the
function $[f:g](x)\eqdef [f(x):g(x)]$ is well defined and is also
called admissible (on $\mathcal{X}$).
Conversely, any $\halfProjLine$-valued function on $\mathcal{X}$ can
be written as $[f:g]$ with $f$ and $g$ two real-valued functions on
$\mathcal{X}$.
Below are some properties of $\halfProjLine$-valued functions.

\begin{proposition}\label{pr:HPL1}
The $(\halfProjLine,\borel(\halfProjLine))$-valued function $[f:g]$ on
$(\mathcal{X},\mathfrak{X})$ is measurable if $f$ and $g$ are both
measurable.
\end{proposition}

\begin{proposition}\label{pr:HPL2}
If $[f_1:g_1]=[f_2:g_2]$ with $f_1$, $f_2$, $g_1$, and $g_2$ all
real-valued measurable functions on $\mathcal{X}$, then there is a
real-valued measurable function $t$ on $\mathcal{X}$ such that
$t(x)\ne 0$, $f_1(x)=t(x)f_2(x)$, and $g_1(x)=t(x)g_2(x)$ for all
$x\in\mathcal{X}$.
\end{proposition}

Let $\mu$ be a measure on $(\mathcal{X},\mathfrak{X})$.
If $[f:g]$ is admissible on $\mathcal{X}$ except a $\mu$-negligible
set of points, then we say $[f:g]$ is admissible $\mu$-almost
everywhere.
Similarly, if $[f_1:g_1]=[f_2:g_2]$ is true for all $x\in\mathcal{X}$
except a $\mu$-negligible set of points, we say $[f_1:g_1]=[f_2:g_2]$
$\mu$-almost everywhere.

\begin{proposition}\label{pr:ConditionalExcpectationOfRatio}
Let $f$ and $g$ be two real-valued integrable functions on the measure
space $(\mathcal{X},\mathfrak{X},\mu)$.
If $[f:g]$ is admissible $\mu$-almost everywhere, then the conditional
expectation
\(
\mu^{\mathfrak{F}}[f:g]\eqdef [\mu^{\mathfrak{F}}f:\mu^{\mathfrak{F}}g]
\)
with respect to some sub $\sigma$-algebra $\mathfrak{F}$ is also
admissible $\mu$-almost everywhere.
\end{proposition}

We are now ready to define the general form of likelihood ratios.

\begin{definition}\label{de:LikelihoodRatio}
Let $\mu$ and $\nu$ be two measures on $(\mathcal{X},\mathfrak{X})$.
The \emph{likelihood ratio} $[\diff\mu:\diff\nu]$ of $\mu$ to $\nu$ is
defined to be $[\diff\mu/\diff(\mu+\nu):\diff\nu/\diff(\mu+\nu)]$,
which is admissible $(\mu+\nu)$-almost everywhere.
\end{definition}

Likelihood ratios enjoy the following property, which is an easy
consequence of Corollary~\ref{co:FiniteApproximation}.

\begin{theorem}\label{th:FiniteApproximation.X}
Let $\mu$ and $\nu$ be two measures on
$(\mathcal{X}\times\mathcal{Y},\mathfrak{X}\times\mathfrak{Y})$.
Let $\xi$ be a measure such that $\xi\ll\mu+\nu$.
Then for any $\epsilon>0$, there is a finite sub $\sigma$-algebra
$\mathfrak{C}$ of $\mathfrak{X}$ and a finite sub $\sigma$-algebra
$\mathfrak{D}$ of $\mathfrak{Y}$ such that
\[
\xi\{\distHPL([\diff\mu|_{\mathfrak{F}}:\diff\nu|_{\mathfrak{F}}],
 [\diff\mu:\diff\nu])\ge\epsilon\}
< \epsilon
\]
for every $\sigma$-algebra satisfying
$\mathfrak{C}\times\mathfrak{D}\subseteq\mathfrak{F}
 \subseteq\mathfrak{X}\times\mathfrak{Y}$.
\end{theorem}

\section{An Example: Separate Random Number Generation from Correlated
 Sources}

In this section, we will explain the quantization approach by an
example: separate random number generation from correlated sources.
For its importance in information theory, the readers are referred to
\cite{yassaee_achievability_2014}.
The finite-alphabet case of this problem has been extensively studied
in \cite{yang_separate_2014} and the references therein.
We will now extend this result to the case of general alphabets.

We first briefly introduce the problem of separate random number
generation.
For simplicity, we only consider the case of two correlated sources
with side information at the tester.

Let $X=(X_0,X_1,X_2)$ be a triple of correlated random elements in
$\mathcal{X}_0\times\mathcal{X}_1\times\mathcal{X}_2$.
Let $\varphi=(\varphi_1,\varphi_2)$ be a pair of (randomness)
extractors $\mathcal{X}_i\to\mathcal{Y}_i$ with $\mathcal{Y}_i$ finite
($i=1,2$).
We are interested in the minimum value of the statistical distance
\[
d(X\mid\varphi)
\eqdef
\statDistance(P_{X_0\varphi_1(X_1)\varphi_2(X_2)},
 P_{X_0}\uniDistribution{\mathcal{Y}_1}
 \uniDistribution{\mathcal{Y}_2})
\]
over all pairs $\varphi$ of extractors, where
$\uniDistribution{\mathcal{Y}_1}$ and
$\uniDistribution{\mathcal{Y}_2}$ denote the uniform distributions
over $\mathcal{Y}_1$ and $\mathcal{Y}_2$, respectively.
The next theorem gives one-shot bounds of $d(X\mid\varphi)$ in the
case of finite alphabets.

\begin{theorem}[\cite{yang_separate_2014}]\label{th:RNG.finite}
Let $X=(X_0,X_1,X_2)$ be a triple of correlated random elements in a
finite product alphabet
$\mathcal{X}_0\times\mathcal{X}_1\times\mathcal{X}_2$.

1) For $r>1$, there exists a pair $\varphi$ of extractors such that
\[
d(X\mid\varphi)
\le P\{(T_X^i(X))_{i=1}^3\notin A_r\} + \frac{\sqrt{3}}{2}r^{-1/2},
\]
where
\begin{IEEEeqnarray*}{rCl}
T_X^1(x_0,x_1,x_2) &\eqdef &\frac{1}{P_{X_1\mid X_0}(x_1\mid x_0)},\\
T_X^2(x_0,x_1,x_2) &\eqdef &\frac{1}{P_{X_2\mid X_0}(x_2\mid x_0)},\\
T_X^3(x_0,x_1,x_2) &\eqdef &\frac{1}{P_{X_1X_2\mid X_0}(x_1,x_2\mid x_0)},
\end{IEEEeqnarray*}
$A_r\eqdef I_{r|\mathcal{Y}_1|}\times I_{r|\mathcal{Y}_2|}
 \times I_{r|\mathcal{Y}_1||\mathcal{Y}_2|}$,
and $I_t\eqdef (t,+\infty)$.

2) Conversely, every pair $\varphi$ of extractors satisfies
\[
d(X\mid\varphi)
\ge P\{(T_X^i(X))_{i=1}^3\notin A_r\} - 3r
\]
for all $0<r<1$.
\end{theorem}

Now let us prove a general-alphabet version of
Theorem~\ref{th:RNG.finite}.

\begin{theorem}\label{th:RNG.Polish}
Let $X=(X_0,X_1,X_2)$ be a triple of correlated random elements in
$(\mathcal{X}_0\times\mathcal{X}_1\times\mathcal{X}_2,
 \mathfrak{X}_0\times\mathfrak{X}_1\times\mathfrak{X}_2)$.%
\footnote{
 It is assumed that the $\sigma$-algebra of a countable alphabet is
 its power set.
}

1) For $r>1$ and $\epsilon>0$, there exists a pair $\varphi$ of
extractors such that
\begin{equation}
d(X\mid\varphi)
\le P\{(T_X^i(X))_{i=1}^3\notin A_r\}
 + \frac{\sqrt{3}}{2}r^{-1/2} + \epsilon,\label{eq:RNG.Polish.a}
\end{equation}
where
\begin{IEEEeqnarray*}{rCl}
T_X^1(x_0,x_1,x_2)
&\eqdef &[\diff P_{X_0X_1X_1}:\diff P_{X_0X_1}P_{X_1\mid X_0}](x_0,x_1,x_1),\\
T_X^2(x_0,x_1,x_2)
&\eqdef &[\diff P_{X_0X_2X_2}:\diff P_{X_0X_2}P_{X_2\mid X_0}](x_0,x_2,x_2),\\
T_X^3(x_0,x_1,x_2)
&\eqdef &[\diff P_{X_0X_1X_2X_1X_2}\\
& &:\diff P_{X_0X_1X_2}P_{X_1X_2\mid X_0}](x_0,x_1,x_2,x_1,x_2),
\end{IEEEeqnarray*}
$A_r\eqdef I_{r|\mathcal{Y}_1|}\times I_{r|\mathcal{Y}_2|}
 \times I_{r|\mathcal{Y}_1||\mathcal{Y}_2|}$,
and $I_t\eqdef ([t:1],[1:0])$.

2) Conversely, every pair $\varphi$ of extractors satisfies
\[
d(X\mid\varphi)
\ge P\{(T_X^i(X))_{i=1}^3\notin A_r\} - 3r
\]
for all $0<r<1$.
\end{theorem}

\begin{IEEEproof}
When the alphabets are all finite, it is clear that
\begin{IEEEeqnarray*}{rCl}
T_X^1(x_0,x_1,x_2)
&= &[1:P_{X_1\mid X_0}(x_1\mid x_0)]\\
T_X^2(x_0,x_1,x_2)
&= &[1:P_{X_2\mid X_0}(x_2\mid x_0)]\\
T_X^3(x_0,x_1,x_2)
&= &[1:P_{X_1X_2\mid X_0}(x_1,x_2\mid x_0)]
\end{IEEEeqnarray*}
$P_X$-almost everywhere, and thus the theorem is trivially true because
of Theorem~\ref{th:RNG.finite}.

1) Direct part:
We first show that the direct part is true for general
$(\mathcal{X}_0,\mathfrak{X}_0)$ and finite $\mathcal{X}_i$ for
$i\in\fromOneTo{2}$.
By Trick~\ref{tr:Approximation} with
Theorems~\ref{th:ConditionalProbability} and
\ref{th:FiniteApproximation.X}, we can find a sequence
$(\mathfrak{F}_n)_{n=1}^\infty$ of finite sub $\sigma$-algebras of
$\mathfrak{X}_0$ such that
\begin{equation}
\lim_{n\to\infty} \expect_{P_{X_0}}\statDistance(P_{X_1X_2\mid X_0=x_0},
 P_{X_1X_2\mid Z_{n,0}=\pi_{\mathfrak{F}_n}(x_0)})
= 0 \label{eq:RNG.Polish.1}
\end{equation}
and
\begin{equation}
\distHPL(T_X^i,T_{g_n(X)}^i\circ g_n)
\tovar{P_X} 0 \label{eq:RNG.Polish.2}
\end{equation}
for $i\in\fromOneTo{3}$, where
$g_n(x_0,x_1,x_2) = (\pi_{\mathfrak{F}_n}(x_0),x_1,x_2)$
and $Z_{n,0}=\pi_{\mathfrak{F}_n}(X_0)$.
By Theorem~\ref{th:RNG.finite}, there is a pair
$\varphi_n=(\varphi_{n,1},\varphi_{n,2})$ of extractors such that
\[
d(g_n(X)\mid\varphi_n)
\le P\{(T_{g_n(X)}^i(g_n(X)))_{i=1}^3 \notin A_r\}
 + \frac{\sqrt{3}}{2}r^{-1/2}
\]
and it follows from Corollary~\ref{co:StatisticalDistance} that
\begin{IEEEeqnarray*}{l}
\bigl|d(X\mid\varphi_n) - d(g_n(X)\mid\varphi_n)\bigr|\\
= \bigl|\expect_{P_{X_0}}
 \statDistance\left(P_{\varphi_{n,1}(X_1)\varphi_{n,2}(X_2)\mid X_0=x_0},
 \uniDistribution{\mathcal{Y}_1}
 \uniDistribution{\mathcal{Y}_2}\right)\\
\quad -\:\expect_{P_{X_0}}
 \statDistance\left(P_{\varphi_{n,1}(X_1)\varphi_{n,2}(X_2)\mid
  Z_{n,0}=\pi_{\mathfrak{F}_n}(x_0)},
 \uniDistribution{\mathcal{Y}_1}
 \uniDistribution{\mathcal{Y}_2}\right)\bigr|\\
\le \expect_{P_{X_0}}\bigl|
 \statDistance\left(P_{\varphi_{n,1}(X_1)\varphi_{n,2}(X_2)\mid X_0=x_0},
 \uniDistribution{\mathcal{Y}_1}
 \uniDistribution{\mathcal{Y}_2}\right)\\
\quad -\:\statDistance\left(P_{\varphi_{n,1}(X_1)\varphi_{n,2}(X_2)\mid
  Z_{n,0}=\pi_{\mathfrak{F}_n}(x_0)},
 \uniDistribution{\mathcal{Y}_1}
 \uniDistribution{\mathcal{Y}_2}\right)\bigr|\\
\le \expect_{P_{X_0}}
 \statDistance\bigl(P_{\varphi_{n,1}(X_1)\varphi_{n,2}(X_2)\mid X_0=x_0},\\
\quad P_{\varphi_{n,1}(X_1)\varphi_{n,2}(X_2)\mid
  Z_{n,0}=\pi_{\mathfrak{F}_n}(x_0)}\bigr)\\
\le \expect_{P_{X_0}}
 \statDistance\bigl(P_{X_1X_2\mid X_0=x_0},P_{X_1X_2\mid
  Z_{n,0}=\pi_{\mathfrak{F}_n}(x_0)}\bigr)
= \littleO(1),
\end{IEEEeqnarray*}
so that
\[
\limsup_{n\to\infty} d(X\mid\varphi_n)
\le P\{(T_{X}^i(X))_{i=1}^3 \notin A_r\}
 + \frac{\sqrt{3}}{2}r^{-1/2}
\]
by the Portmanteau theorem
\cite[Theorem~3.25]{kallenberg_foundations_1997} with
\eqref{eq:RNG.Polish.2}, and therefore $\varphi_n$ satisfies
\eqref{eq:RNG.Polish.a} for sufficiently large $n$.

We are now ready to prove the general case.
By Trick~\ref{tr:Approximation} with
Theorem~\ref{th:FiniteApproximation.X}, we can find a sequence
$((\mathfrak{C}_{n,1},\mathfrak{C}_{n,2}))_{n=1}^\infty$ of pairs of
finite sub $\sigma$-algebras such that, for all $i\in\fromOneTo{3}$,
\begin{equation}
\distHPL(T_{X}^i,T_{h_n(X)}^i\circ h_n)
\tovar{P_X} 0, \label{eq:RNG.Polish.3}
\end{equation}
where
$h_n(x_0,x_1,x_2) = (x_0,\pi_{\mathfrak{C}_{n,1}}(x_1),
 \pi_{\mathfrak{C}_{n,2}}(x_2))$.
Then, for every $n$, there is a pair
$\psi_n=(\psi_{n,1},\psi_{n,2})$ of extractors such that
\begin{IEEEeqnarray*}{rCl}
d(h_n(X)\mid\psi_n)
&\le  &P\{(T_{h_n(X)}^i(h_n(X)))_{i=1}^3\notin A_r\}\\
& &+\:\frac{\sqrt{3}}{2}r^{-1/2}+\frac{\epsilon}{2}.
\end{IEEEeqnarray*}
Let
$\varphi_n
 = (\psi_{n,1}\circ\pi_{\mathfrak{C}_{n,1}},
  \psi_{n,2}\circ\pi_{\mathfrak{C}_{n,2}})$.
We further have
\begin{IEEEeqnarray*}{l}
\limsup_{n\to\infty} d(X\mid\varphi_n)\\
= \limsup_{n\to\infty} d(h_n(X)\mid\psi_n)\\
\le \limsup_{n\to\infty}
 P\{(T_{h_n(X)}^i(h_n(X)))_{i=1}^3\notin A_r\}
 + \frac{\sqrt{3}}{2}r^{-1/2} + \frac{\epsilon}{2}\\
\le  P\{(T_{X}^i(X))_{i=1}^3\notin A_r\} + \frac{\sqrt{3}}{2}r^{-1/2}
 + \frac{\epsilon}{2},
\end{IEEEeqnarray*}
where the last inequality follows from \eqref{eq:RNG.Polish.3} and the
Portmanteau theorem, and therefore $\varphi_n$ satisfies
\eqref{eq:RNG.Polish.a} for sufficiently large $n$.

2) Converse part:
Similar to the proof of direct part, we first prove the converse part
with general $(\mathcal{X}_0,\mathfrak{X}_0)$ and other alphabets finite.
Again by Trick~\ref{tr:Approximation} with
Theorems~\ref{th:ConditionalProbability} and
\ref{th:FiniteApproximation.X}, we have \eqref{eq:RNG.Polish.1}
and \eqref{eq:RNG.Polish.2}, so that
\begin{IEEEeqnarray*}{rCl}
d(X\mid\varphi)
&= &\lim_{n\to\infty} d(g_n(X)\mid\varphi)\\
&\ge &\liminf_{n\to\infty}
 P\{(T_{g_n(X)}^i(g_n(X)))_{i=1}^3\notin\closure{A_r}\}-3r\\
&\ge & P\{(T_X^i(X))_{i=1}^3\notin\closure{A_r}\}-3r,
\end{IEEEeqnarray*}
where the last inequality follows from the Portmanteau theorem, and
$\closure{A_r}$ denotes the closure of $A_r$.

Then we turn to the general case.
By Trick~\ref{tr:Approximation} with
Theorem~\ref{th:FiniteApproximation.X}, we can find a sequence
$((\mathfrak{C}_{n,1},\mathfrak{C}_{n,2}))_{n=1}^\infty$ of pairs of
finite sub $\sigma$-algebras satisfying \eqref{eq:RNG.Polish.3} and
$\sigma(\varphi_i)\subseteq\mathfrak{C}_{n,i}$ for all $n$ and $i$.
Then for every $n$ and $i$, we have
$\varphi_i=\psi_{n,i}\circ\pi_{\mathfrak{C}_{n,i}}$ for some
$\psi_{n,i}$, and therefore
\begin{IEEEeqnarray*}{rCl}
d(X\mid\varphi)
&= &\lim_{n\to\infty} d(h_n(X)\mid\psi_n)\\
&\ge &\liminf_{n\to\infty}
 P\{(T_{g_n(X)}^i(g_n(X)))_{i=1}^3\notin\closure{A_r}\}-3r\\
&\ge &P\{(T_X^i(X))_{i=1}^3\notin\closure{A_r}\}-3r,
\end{IEEEeqnarray*}
where the last inequality follows from the Portmanteau theorem.
Finally,
\begin{IEEEeqnarray*}{rCl}
d(X\mid\varphi)
&\ge &\lim_{k\to\infty} \left(
 P\{(T_X^i(X))_{i=1}^3\notin\closure{A_{s_k}}\}-3s_k \right)\\
&\ge &P\left\{(T_X^i(X))_{i=1}^3\notin
 \bigcup_{k=1}^\infty\closure{A_{s_k}}\right\}-3r\\
&= &P\{(T_X^i(X))_{i=1}^3\notin A_r\}-3r,
\end{IEEEeqnarray*}
where $s_k=(r+1/k)\wedge 1$, namely, the minimum of $r+1/k$ and $1$.
\end{IEEEproof}

\begin{remark}
Note that even if the alphabets $\mathcal{X}_i$ are all continuous,
there are many nontrivial situations in which $T_X^i(X)$ does not
degenerate to the point $[1:0]$.
For simplicity, let us assume that $X_2$ is constant.
If $X_1$ is a $(p,1-p)$-mixture of a real number and a uniform random
variable in $[0,1]$, both independent of $X_0$, then $T_X^3(X)$ is a
random variable taking values in $\{1/p,[1:0]\}$ with probabilities
$p$ and $1-p$, respectively.
Note that the extractors in our problem are fixed-length extractors.
This example tells us that the performance of a fixed-length extractor
is dominated by the worst case.
If $X_1$ takes values in $\{X_0-0.5,X_0+0.5\}$ with equal
probabilities and $X_0$ is uniformly distributed over $[0,1]$, then
$T_X^3(X)=2$ almost surely.
This example shows that a random variable even with continuous
distribution does not necessarily have infinite randomness because of
the side information at the tester.

The one-shot bounds provided by Theorem~\ref{th:RNG.Polish} are tight
enough for the first- and second-order asymptotic analysis.
If we define
\[
\ln[x:y]\eqdef\ln(x)-\ln(y)\in[-\infty,+\infty]
\]
(which is well defined because $(x,y)\ne (0,0)$), then we can easily
obtain a generalization of the achievable rate region in the finite
case, with for example the \term{spectral inf-entropy rate}
$\infEntropyRate(X_1\mid X_0)$ of general source $X_1$ given general
source $X_0$ defined by
\[
\pliminf_{n\to\infty} \frac{1}{n} \ln
 \big[P_{X_0^{(n)}X_1^{(n)}X_1^{(n)}}
 :P_{X_0^{(n)}X_1^{(n)}}P_{X_1^{(n)}\mid X_0^{(n)}}\big].
\]
\end{remark}

\omitted{\section{The Proofs of Results in Section~\ref{sec:Quantization}}
\label{sec:Proof}

\begin{IEEEproof}[Proof of Corollary~\ref{co:FiniteApproximation}]
Since $\nu\ll\mu$, it follows from
\cite[Lemma~4.2.1]{cohn_measure_2013} that there is a positive
$\delta$ such that each measurable set $A$ satisfying $\mu(A)<\delta$
also satisfies $\nu(A)<\epsilon$.
Using Theorem~\ref{th:FiniteApproximation} with
$\epsilon'=\delta\epsilon/2$,
we obtain by Markov's inequality that there is a finite sub
$\sigma$-algebra $\mathfrak{C}$ of $\mathfrak{X}$ and a finite sub
$\sigma$-algebra $\mathfrak{D}$ of $\mathfrak{Y}$ such that
\[
\mu\{|\mu^{\mathfrak{F}}f-f|\ge\epsilon\}
\le \frac{\mu|\mu^{\mathfrak{F}}f-f|}{\epsilon}
= \frac{\delta}{2}
\]
for every $\sigma$-algebra $\mathfrak{F}$ satisfying
$\mathfrak{C}\times\mathfrak{D}\subseteq\mathfrak{F}
 \subseteq\mathfrak{X}\times\mathfrak{Y}$,
so that $\nu\{|\mu^{\mathfrak{F}}f-f|\ge\epsilon\}<\epsilon$.
\end{IEEEproof}

\begin{IEEEproof}[Proof of Theorem~\ref{th:ConditionalProbability}]
Since $\expect_\lambda(\mu)$ is a probability measure on
$\mathcal{Y}$, there is a finite subset $B$ of $\mathcal{Y}$ such that
\[
(\expect_\lambda(\mu))(B^\setComplement)
=\expect_\lambda(\mu^{B^\setComplement})
<\frac{\epsilon}{2}.
\]
For every $y\in B$, it follows from
Theorem~\ref{th:FiniteApproximation} that there is a finite sub
$\sigma$-algebra $\mathfrak{C}_y$ of $\mathfrak{X}$ such that
\[
\expect_\lambda
 \left|\expect_\lambda^{\mathfrak{F}}\mu^{\{y\}}-\mu^{\{y\}}\right|
< \frac{\epsilon}{|B|}.
\]
for every $\sigma$-algebra $\mathfrak{F}$ satisfying
$\mathfrak{C}_y\subseteq\mathfrak{F}\subseteq\mathfrak{X}$.
Let $\mathfrak{C}=\sigma(\bigcup_{y\in B}\mathfrak{C}_y)$, and then
for every $\sigma$-algebra $\mathfrak{F}$ satisfying
$\mathfrak{C}\subseteq\mathfrak{F}\subseteq\mathfrak{X}$,
\begin{IEEEeqnarray*}{l}
\expect_\lambda\statDistance(\expect_\lambda^\mathfrak{F}\mu,\mu)\\
= \int \frac{1}{2} \sum_{y\in\mathcal{Y}}
 |\expect_\lambda^\mathfrak{F}\mu(x,\{y\})-\mu(x,\{y\})|
 \lambda(\diff x)\\
\le \int \frac{1}{2} \sum_{y\in B}
 |\expect_\lambda^\mathfrak{F}\mu(x,\{y\})-\mu(x,\{y\})|
 \lambda(\diff x)\\
\quad +\:\int \frac{1}{2} (
 \expect_\lambda^\mathfrak{F}\mu(x,B^\setComplement)
 + \mu(x,B^\setComplement)) \lambda(\diff x)\\
= \frac{1}{2} \sum_{y\in B} \expect_\lambda
 \left|\expect_\lambda^\mathfrak{F}\mu^{\{y\}}-\mu^{\{y\}}\right|
 + \expect_\lambda(\mu^{B^\setComplement})
< \epsilon.
\end{IEEEeqnarray*}
\end{IEEEproof}

\begin{IEEEproof}[Proof of Theorem~\ref{th:KernelRatioToMeasureRatio}]
By Propositions~\ref{pr:KernelProperty1} and \ref{pr:KernelProperty2},
we immediately have $\lambda\mu\ll\lambda\nu$.
For any $C\in\mathfrak{X}\times\mathfrak{Y}$,
\begin{eqnarray*}
\int \mu_x(C_x) \lambda(\diff x)
&= &\int \overline{\mu}_x(C) \lambda(\diff x)\\
&= &(\lambda\mu)(C)\\
&= &\int_C \frac{\diff(\lambda\mu)}{\diff(\lambda\nu)}
 \diff(\lambda\nu)\\
&= &\int \lambda(\diff x) \int_C \frac{\diff(\lambda\mu)}
 {\diff(\lambda\nu)} \diff\overline{\nu}_x\\
&= &\int \lambda(\diff x) \int_{C_x} \frac{\diff(\lambda\mu)}
 {\diff(\lambda\nu)}(x,y) \nu_x(\diff y),
\end{eqnarray*}
where the last equality follows from
Proposition~\ref{pr:KernelProperty3}.
Taking $C=A\times B$ for any $A\in\mathfrak{X}$ and
$B\in\mathfrak{Y}$, we thus have
\[
\int_A \mu_x(B) \lambda(\diff x)
= \int_A \lambda(\diff x) \int_B \frac{\diff(\lambda\mu)}
 {\diff(\lambda\nu)}(x,y) \nu_x(\diff y),
\]
so that
\[
\mu_x(B)
= \int_B \frac{\diff(\lambda\mu)}{\diff(\lambda\nu)}(x,y)
 \nu_x(\diff y),
\]
for $\lambda$-almost every $x$ in $\mathcal{X}$, and for every such
$x$,
\[
\frac{\diff\mu_x}{\diff\nu_x}(y)
= \frac{\diff(\lambda\mu)}{\diff(\lambda\nu)}(x,y)
\]
for $\nu_x$-almost every $y$ in $\mathcal{Y}$.
If $\diff\mu_x/\diff\nu_x$ has a
$(\mathfrak{X}\times\mathfrak{Y})$-measurable version $f$, then
\[
f=\frac{\diff(\lambda\mu)}{\diff(\lambda\nu)}
\]
for all $(x,y)$ except a $\lambda\nu$-negligible set of points
(Proposition~\ref{pr:KernelProperty4}).
\end{IEEEproof}

\begin{IEEEproof}[Proof of Corollary~\ref{co:StatisticalDistance}]
\begin{IEEEeqnarray}{l}
\statDistance(\lambda\mu,\lambda\nu) \nonumber\\
= \frac{1}{2} \int
 \left|\frac{\diff\lambda\mu}{\diff(\lambda\kp(\mu+\nu))}
 - \frac{\diff\lambda\nu}{\diff(\lambda\kp(\mu+\nu))}\right|
 \diff(\lambda\kp(\mu+\nu)) \nonumber\\
= \frac{1}{2} \int \lambda(\diff x)
 \int \bigg|\frac{\diff\lambda\mu}{\diff(\lambda\kp(\mu+\nu))}(x,y)
 \nonumber\\
\quad {}-\frac{\diff\lambda\nu}{\diff(\lambda\kp(\mu+\nu))}(x,y)\bigg|
 (\overline{\mu+\nu})_x(\diff(x,y)) \nonumber\\
= \frac{1}{2} \int \lambda(\diff x)
 \int \left|\frac{\diff\mu_x}{\diff(\mu+\nu)_x}
 - \frac{\diff\nu_x}{\diff(\mu+\nu)_x}\right| \diff(\mu+\nu)_x
 \label{eq:KernelProperty5.1}\\
= \int \statDistance(\mu_x,\nu_x) \lambda(\diff x)
= \expect_\lambda \statDistance(\mu_x,\nu_x),\nonumber
\end{IEEEeqnarray}
where \eqref{eq:KernelProperty5.1} follow from
Theorem~\ref{th:KernelRatioToMeasureRatio} and
Proposition~\ref{pr:KernelProperty3}.
\end{IEEEproof}

\begin{IEEEproof}[Proof of Proposition~\ref{pr:HPL1}]
It is obvious by observing that $[f:g]=\pi_\halfProjLine(f(x),g(x))$
with $f$, $g$, and $\pi_\halfProjLine$ all measurable.
\end{IEEEproof}

\begin{IEEEproof}[Proof of Proposition~\ref{pr:HPL2}]
By definition, for every $x\in\mathcal{X}$, there is a number
$t(x)\ne 0$ such that $f_1(x)=t(x)f_2(x)$ and $g_1(x)=t(x)g_2(x)$.
Then it suffices to show that $t$ is measurable.
Let $A=\{x:f_2(x)\ne 0\}$.
It is clear that
\[
t = \frac{f_1}{f_21_A+1_{A^\setComplement}}1_A
 + \frac{g_1}{g_21_{A^\setComplement}+1_A} 1_{A^\setComplement},
\]
which is measurable.
\end{IEEEproof}

\begin{IEEEproof}[Proof of Proposition~\ref{pr:ConditionalExcpectationOfRatio}]
Since $[f:g]$ is admissible $\mu$-almost everywhere, we have $f\ge 0$,
$g\ge 0$, and $(f,g)\ne (0,0)$ $\mu$-almost everywhere, so that
$\mu^{\mathfrak{F}}f\ge 0$, $\mu^{\mathfrak{F}}g\ge 0$, and
$(\mu^{\mathfrak{F}}f,\mu^{\mathfrak{F}}g)\ne (0,0)$ $\mu$-almost
everywhere, and therefore $\mu^{\mathfrak{F}}[f:g]$ is admissible
$\mu$-almost everywhere.
\end{IEEEproof}

\begin{IEEEproof}[Proof of Theorem~\ref{th:FiniteApproximation.X}]
First note that
\[
[\diff\mu|_{\mathfrak{F}}:\diff\nu|_{\mathfrak{F}}]
= (\mu+\nu)^{\mathfrak{F}}[\diff\mu:\diff\nu],
\]
which is admissible $(\mu+\nu)$-almost everywhere by
Proposition~\ref{pr:ConditionalExcpectationOfRatio}.
From Corollary~\ref{co:FiniteApproximation}, it follows that there is
a finite sub $\sigma$-algebra $\mathfrak{C}$ of $\mathfrak{X}$ and a
finite sub $\sigma$-algebra $\mathfrak{D}$ of $\mathfrak{Y}$ such that
\begin{IEEEeqnarray*}{l}
\xi\left\{\left|
 \frac{\diff\mu|_{\mathfrak{F}}}{\diff(\mu+\nu)|_{\mathfrak{F}}}
 -\frac{\diff\mu}{\diff(\mu+\nu)}
 \right|\ge\epsilon\right\}\\
\qquad = \xi\left\{\left|
 (\mu+\nu)^{\mathfrak{F}}\frac{\diff\mu}{\diff(\mu+\nu)}
 -\frac{\diff\mu}{\diff(\mu+\nu)}
 \right|\ge\epsilon\right\}
< \epsilon
\end{IEEEeqnarray*}
for every $\sigma$-algebra $\mathfrak{F}$ satisfying
$\mathfrak{C}\times\mathfrak{D}\subseteq\mathfrak{F}
 \subseteq\mathfrak{X}\times\mathfrak{Y}$.
Therefore
\begin{IEEEeqnarray*}{l}
\xi\{\distHPL([\diff\mu|_{\mathfrak{F}}:\diff\nu|_{\mathfrak{F}}],
 [\diff\mu:\diff\nu])\ge\epsilon\}\\
\qquad = \xi\left\{\left|
 \frac{\diff\mu|_{\mathfrak{F}}}{\diff(\mu+\nu)|_{\mathfrak{F}}}
 -\frac{\diff\mu}{\diff(\mu+\nu)}
 \right|\ge\epsilon\right\}
< \epsilon.
\end{IEEEeqnarray*}
\end{IEEEproof}

\section{Facts Used By Section~\ref{sec:Proof}}

\begin{proposition}\label{pr:KernelProperty1}
Let $\mu$ and $\nu$ be two kernels from the measure space
$(\mathcal{X},\mathfrak{X},\lambda)$ to $(\mathcal{Y},\mathfrak{Y})$
such that $\mu_x\ll\nu_x$ for $\lambda$-almost every $x$ in
$\mathcal{X}$.
Then $\lambda(\mu)\ll\lambda(\nu)$.
\end{proposition}

\begin{IEEEproof}
Let $B\in\mathfrak{Y}$.
If $\lambda(\nu^B)=0$, then $\nu(x,B)=0$ for
$\lambda$-almost every $x$ in $\mathcal{X}$, so that $\mu(x,B)=0$ for
$\lambda$-almost every $x$ in $\mathcal{X}$, hence $\lambda(\mu^B)=0$,
and therefore $\lambda(\mu)\ll\lambda(\nu)$.
\end{IEEEproof}

\begin{proposition}\label{pr:KernelProperty2}
Let $\mu$ and $\nu$ be two kernels from $(\mathcal{X},\mathfrak{X})$
to $(\mathcal{Y},\mathfrak{Y})$.
For every $x\in\mathcal{X}$, $\mu_x\ll\nu_x$ iff
$\overline{\mu}_x\ll\overline{\nu}_x$.
\end{proposition}

\begin{IEEEproof}
Let $B\in\mathfrak{Y}$ and $C\in\mathfrak{X}\times\mathfrak{Y}$.

($\Rightarrow$)
If $\overline{\nu}(x,C)=0$, then $\nu(x,C_x)=0$, so that
$\mu(x,C_x)=0$, hence $\overline{\mu}(x,C)=0$, and therefore
$\overline{\mu}_x\ll\overline{\nu}_x$.

($\Leftarrow$)
If $\nu(x,B)=0$, then $\overline{\nu}(x,\mathcal{X}\times B)=0$, so
that $\overline{\mu}(x,\mathcal{X}\times B)=0$, hence $\mu(x,B)=0$,
and therefore $\mu_x\ll\nu_x$.
\end{IEEEproof}

\begin{proposition}\label{pr:KernelProperty3}
Let $f$ be a real-valued measurable function on
$(\mathcal{X}\times\mathcal{Y},\mathfrak{X}\times\mathfrak{Y})$.
Then $\overline{\mu}_xf = \mu_xf_x$.
\end{proposition}

\begin{IEEEproof}
Let $\iota_x$ be the map of $\mathcal{Y}$ into
$\mathcal{X}\times\mathcal{Y}$ given by $y\mapsto (x,y)$, which is
clearly measurable.
Then
\[
\overline{\mu}_x(f)
= (\mu_x \circ \iota_x^{-1})f
= \mu_x (f\circ \iota_x)
= \mu_x f_x,
\]
where the second equality follows from
\cite[Lemma~2.6.8]{cohn_measure_2013}.
\end{IEEEproof}

\begin{proposition}\label{pr:KernelProperty4}
Let $\mu$ be a kernel from the measure space
$(\mathcal{X},\mathfrak{X},\lambda)$ to $(\mathcal{Y},\mathfrak{Y})$.
Let $A\in\mathfrak{X}$ and $C\in\mathfrak{X}\times\mathfrak{Y}$.
If $\lambda(A)=0$ and $\mu(x,C_x)=0$ for all $x\notin A$, then
$(\lambda\mu)(C)=0$.
\end{proposition}

\begin{IEEEproof}
\begin{IEEEeqnarray*}{rCl}
(\lambda\mu)(C)
&= &(\lambda\mu)(C\cap (A^\setComplement\times\mathcal{Y}))
 + (\lambda\mu)(C\cap (A\times\mathcal{Y}))\\
&\le &\int \overline{\mu}(x,C\cap (A^\setComplement\times\mathcal{Y}))
 \lambda(\diff x) + (\lambda\mu)(A\times\mathcal{Y})\\
&= &\int_{A^\setComplement} \mu(x,C_x) \lambda(\diff x) + \lambda(A)
= 0.
\end{IEEEeqnarray*}
\end{IEEEproof}
}

\section{Conclusion}

In this paper, we develop a general approach for deriving one-shot
bounds for information-theoretic problems on general alphabets.
This approach provides a mechanical way for solving problems on
general alphabets based on their solutions in the finite-alphabet
case, and hence it helps us better understand information theory in a
unified way beyond countable alphabets.
This is still an ongoing research.
Applying this approach to other problems of information theory will be
our future work.


\section*{Acknowledgment}

This work was supported in part by the National Natural
Science Foundation of China under Grant 61571398, Grant 61571006, and
Grant 61371094,
in part by the National Key Basic Research Program of China under
Grant 2012CB316104,
in part by the Zhejiang Provincial Natural Science Foundation under
Grant LR12F01002,
and in part by the open project of Zhejiang Provincial Key Laboratory
of Information Processing, Communication and Networking.



\bibliographystyle{IEEEtran}
\bibliography{IEEEabrv,urng2}
%

\small\raggedleft\textcolor{red}{\texttt{(Version~\docVersion.\docBuildNumber)}}
\end{document}